\documentclass{article}

\usepackage{PRIMEarxiv}

\usepackage[utf8]{inputenc} 
\usepackage[T1]{fontenc}    
\usepackage{hyperref}       
\usepackage{url}            
\usepackage{booktabs}       
\usepackage{amsfonts}       
\usepackage{nicefrac}       
\usepackage{microtype}      
\usepackage{lipsum}
\usepackage{fancyhdr}       
\usepackage{graphicx}       
\graphicspath{{media/}}     

\title{An Artificial Neuron for Enhanced Problem Solving in Large Language Models
}

\author{
  Sumedh Rasal \\
  Georgia Institute of Technology \\
  Chicago, IL\\
  \texttt{srasal3@gatech.edu} \\
}

\begin{document}
\maketitle

\begin{abstract}
Recent advancements in artificial intelligence have propelled the capabilities of Large Language Models (LLMs), yet their ability to mimic nuanced human reasoning remains limited. This paper introduces a novel conceptual enhancement to LLMs, termed the "Artificial Neuron," designed to significantly bolster cognitive processing by integrating external memory systems. This enhancement mimics neurobiological processes, facilitating advanced reasoning and learning through a dynamic feedback loop mechanism. We propose a unique framework wherein each LLM interaction—specifically in solving complex math word problems and common sense reasoning tasks—is recorded and analyzed. Incorrect responses are refined using a higher-capacity LLM or human-in-the-loop corrections, and both the query and the enhanced response are stored in a vector database, structured much like neuronal synaptic connections. This "Artificial Neuron" thus serves as an external memory aid, allowing the LLM to reference past interactions and apply learned reasoning strategies to new problems.

Our experimental setup involves training with the GSM8K dataset for initial model response generation, followed by systematic refinements through feedback loops. Subsequent testing demonstrated a 15\% improvement in accuracy and efficiency, underscoring the potential of external memory systems to advance LLMs beyond current limitations. This approach not only enhances the LLM's problem-solving precision but also reduces computational redundancy, paving the way for more sophisticated applications of artificial intelligence in cognitive tasks. This paper details the methodology, implementation, and implications of the Artificial Neuron model, offering a transformative perspective on enhancing machine intelligence.

\end{abstract}


\section{Introduction}
The advent of artificial intelligence has ushered in an era where Large Language Models (LLMs) like GPT-2 through GPT-4 \cite{openai2023gpt} \cite{bubeck2023sparks} have demonstrated unprecedented capabilities in language understanding and problem-solving \cite{brown2020language} \cite{wei2022emergent} \cite{touvron2023llama} \cite{devlin2018bert} \cite{patil2023gorilla} \cite{yao2022react} \cite{li2023camel}. These models have progressively bridged the gap between computational text processing and genuine linguistic comprehension, enhancing their utility across a spectrum of applications, from simple task automation to complex problem resolution \cite{chen2021evaluating} \cite{ouyang2022training}. Despite these advancements, the quest towards achieving a fully autonomous LLM, capable of exhibiting human-like reasoning and decision-making, remains elusive \cite{rajani2019explain}.

This paper introduces the concept of the "Artificial Neuron" as a strategic enhancement for LLMs, aiming to replicate aspects of human cognitive processes through the integration of an external memory module. Traditional LLMs, while proficient in language tasks, cannot often draw on past experiences or adapt their responses based on previously encountered contexts. The Artificial Neuron is designed to address these shortcomings by enabling LLMs to store, recall, and utilize past interactions in their decision-making processes, much like the synaptic activities in the human brain \cite{susskind2000using} \cite{zhou2023cast} \cite{luppi2022synergistic}.

In exploring whether we can endow LLMs with a form of artificial wisdom, this study seeks to define and operationalize wisdom within the realm of artificial intelligence. Wisdom, in this context, extends beyond mere data retrieval to encompass the nuanced application of knowledge and experience in problem-solving. The flexibility inherent in the transformer architecture of LLMs allows for linguistic adaptability, yet it does not inherently equip these models with the depth of understanding necessary to discern the subtleties of language usage or to anticipate the implications of newly acquired information.

The introduction of an Artificial Neuron aims to revolutionize this landscape by providing a mechanism through which LLMs can enhance their reasoning skills \cite{ling2017program} \cite{cobbe2021training} \cite{roy2016solving} \cite{chiang2018semantically} \cite{amini2019mathqa} \cite{chen2019neural} \cite{talmor2018commonsenseqa} \cite{sap2019socialiqa} \cite{patel2021nlp}. This paper will detail a novel approach to supplementing LLMs with external memory, thus enabling them to learn from both their training data and their interactive experiences. By employing past encounters as templates for current problem-solving, the Artificial Neuron seeks to improve the accuracy and relevance of LLM responses, ultimately bridging the gap between current capabilities and the potential for true artificial general intelligence \cite{chiang2023can} \cite{rasal2024navigating} \cite{Agarwal2024FaithfulnessVP}.

Our proposed methodology deviates from traditional training methods, focusing instead on building an external memory that aids the LLM in addressing complex math word problems and common sense reasoning challenges. Initially, we collect responses from the LLM using datasets comprised of math problems and common sense questions, documenting both the answers and the LLM's rationale in a vector database. Incorrect responses are identified and refined either through further consultations with the GPT-4 model or via human intervention, ensuring that only corrected answers are stored.

Upon completion of these revisions, the enriched external memory is integrated into the LLM's operational framework. When the LLM encounters a new problem, it performs a semantic search within the vector database to identify the top three similar past scenarios. These records, detailing the strategies previously employed, guide the LLM in formulating its approach to the new question, utilizing a chain-of-thought prompt style based on the most relevant templates.

Adopting a "learning by cases" strategy, combined with chain-of-thought reasoning, we enhance the capabilities of a GPT-3.5-turbo model by equipping it with this strategic external memory. The resultant outputs from this augmented version of the GPT-3.5-turbo are promising and suggest significant improvements over its predecessors. Our approach outlines:
\begin{itemize}
    \item The utilization of a vector database to meticulously record actions performed by the LLM.
    \item A mechanism to evaluate and amend these records, ensuring that they not only reflect successful outcomes but also incorporate corrections where necessary.
    \item A utility function that leverages these enriched records to set contextually appropriate prompts, thereby elevating the LLM's baseline accuracy and performance.
\end{itemize}

This framework not only enhances the accuracy and efficiency of LLMs but also bridges the gap towards achieving a more nuanced, context-aware artificial intelligence.

\section{Methodology}
The methodology presented in this paper revolves around the innovative concept of the "Artificial Neuron" for Large Language Models (LLMs), a structured external memory system designed to enhance the reasoning abilities and contextual understanding of AI in complex problem-solving scenarios. The approach is delineated into distinct phases: setup, interaction, error correction, and performance enhancement through a continuous learning cycle.

\textbf{Phase 1: Setup and Integration of the Artificial Neuron}

\textbf{Model Configuration}: We begin by configuring the LLM with an initial architecture that includes an interface for integrating external memory components. This setup allows the model to access and interact with the Artificial Neuron, a specialized vector database designed to mimic synaptic connections by storing and retrieving problem-solving experiences.

\textbf{External Memory Initialization}: The Artificial Neuron is populated initially with a broad range of problem-solving scenarios drawn from selected datasets, including GSM8K for math word problems and other datasets for common sense reasoning tasks. Each entry in this memory unit contains the problem statement, the LLM’s response, the reasoning process, and metadata about the task.

\textbf{Phase 2: Interactive Learning and Memory Utilization}

\textbf{Data Collection}: As the LLM engages with new problem-solving tasks, each interaction—question, proposed solution, and reasoning pathway—is logged. This continuous data collection is critical for dynamically updating the Artificial Neuron.

\textbf{Semantic Search Mechanism}: When presented with a new problem, the LLM performs a semantic search within the Artificial Neuron to retrieve the most relevant past interactions. This process is facilitated by advanced algorithms that analyze the similarity between the new problem and stored entries, focusing on both the problem’s nature and the contextual nuances of previous solutions.

\textbf{Phase 3: Error Correction and Feedback Integration}

\textbf{Performance Analysis}: Responses generated by the LLM are evaluated for accuracy against established benchmarks or through expert review. Errors identified in the LLM's reasoning or outputs trigger a corrective feedback loop.

\textbf{Feedback Mechanism}: Incorrect responses are corrected either through further processing by a more advanced LLM model or via human intervention. The corrected response, along with an analysis of the error and a detailed explanation of the correct reasoning process, is then fed back into the Artificial Neuron.

\textbf{Memory Update}: The Artificial Neuron is updated with this new information, enhancing its database with both the corrected response and a revised reasoning pathway. This update not only rectifies errors but also enriches the future problem-solving capacity of the LLM.

\textbf{Phase 4: Continuous Improvement and Testing}

\textbf{Iterative Learning}: The process of interaction, error correction, and memory enhancement is cyclic, with the LLM continuously refining its problem-solving strategies based on accumulated experience stored in the Artificial Neuron.

\textbf{Systematic Evaluation}: Regularly scheduled evaluations measure the LLM’s performance improvements over time. These assessments help fine-tune the semantic search algorithms and the effectiveness of the feedback mechanisms, ensuring that the Artificial Neuron remains a robust and dynamic aid in the LLM's cognitive processing.

This methodology not only proposes a practical framework for enhancing the capabilities of LLMs through external memory but also illustrates a pathway towards more sophisticated AI systems capable of learning and adapting from their interactions, much like human cognitive processes.

\section{Experiments}

\subsection{Enhancing Mathematical Reasoning with the Artificial Neuron}
\textbf{Datasets Utilized}

1. \textbf{MAWPS}: An extensive online repository of math word problems designed to serve as a unified platform for evaluating algorithmic performance. This dataset contains 3,320 problems sourced from previous studies, equipped with tools that allow for the customization of datasets by adjusting lexical and template overlaps, as well as filtering out ungrammatical problems. The dynamic, online nature of MAWPS encourages continuous community contributions, making it a rich resource for diverse mathematical challenges \cite{cobbe2021gsm8k}.

2. \textbf{SVAMP}: A challenge set tailored for elementary-level math word problems, presenting simple variations on arithmetic problems encapsulated in natural language narratives. These problems test the model’s ability to interpret and solve questions about unknown quantities based on described scenarios.

\textbf{Training and Testing Procedure}
- For both datasets, half of the problems were allocated for training the Artificial Neuron, while the other half served as a test set to evaluate the enhancement brought about by this novel approach.

- During training, the Artificial Neuron was exposed to various problem-solving scenarios where each interaction was logged, and incorrect responses were subsequently corrected and enriched with detailed feedback.

- In the testing phase, the LLM equipped with the Artificial Neuron tackled new, unseen problems from the test set. The model's responses were compared against a control group, which consisted of the same LLM architecture without the integration of the Artificial Neuron.

\textbf{Results}
- The experiments demonstrated a notable improvement of approximately 15\% in solving accuracy over the baseline GPT-3.5-turbo model. This enhancement underscores the efficacy of the Artificial Neuron in providing contextual guidance and memory-based reasoning to tackle diverse and complex math problems.

\subsection{Complex Sequential Question Answering (CSQA)}
\textbf{Dataset Utilized}

1. \textbf{CSQA Dataset}: A newly introduced dataset designed to simulate real-world interactions involving both question answering (QA) and dialogue systems. The dataset comprises approximately 200,000 dialogues with a total of 1.6 million turns. These dialogues are built around a large-scale knowledge graph (KG) and require advanced inferencing to answer questions that often involve logical, quantitative, and comparative reasoning.

\textbf{Training and Testing Procedure}
- Similarly, half of the CSQA dataset was used to train the Artificial Neuron, focusing particularly on enhancing the model’s ability to parse complex natural language queries, manage conversational context, and utilize KGs effectively.

- The remaining half of the dataset was used to evaluate the model's performance in a sequential question-answering setup that mirrors practical conversational scenarios encountered by modern chatbots.

\textbf{Results}
- Preliminary results indicated that existing state-of-the-art dialogue and QA models were inadequate in handling the complexity of the CSQA dataset. However, the introduction of the Artificial Neuron significantly improved the LLM's ability to engage in coherent dialogue and answer questions based on complex inferencing.

- The experiment revealed an improvement in the model’s capability to clarify ambiguities, resolve coreferences, and retrieve relevant subgraphs from the KG for accurate responses.

These experiments validate the Artificial Neuron’s potential to substantially enhance LLMs' problem-solving capabilities across different domains. By integrating experiences and learned strategies into an external memory, the LLMs exhibited improved reasoning, adaptability, and accuracy in handling complex, real-world tasks. This promising advancement invites further exploration into extending these methods to broader AI applications, potentially paving the way for more sophisticated cognitive abilities in artificial intelligence systems.



\section{Limitations}
While the introduction of the Artificial Neuron represents a significant advancement in the capabilities of Large Language Models (LLMs), this approach is not without its limitations. Traditional methods for enhancing LLM accuracy typically involve extensive retraining \cite{rasal2023beyond} on large datasets, which, while effective, can be prohibitively expensive and time-consuming. Our methodology circumvents these costs by enabling LLMs to learn from interactive experiences stored within an external memory framework, reducing the need for frequent retraining. However, this system introduces new challenges that must be addressed:

\textbf{Dependency on Manual Input}: The current implementation relies heavily on manual interventions for error identification and correction. Each instance where the LLM fails to provide the correct answer or reasoning must be manually reviewed and corrected, which is not only labor-intensive but also limits the scalability of this approach.
   
\textbf{Memory Integration}: While integrating external memory with an LLM seems straightforward, optimizing this memory for efficient retrieval and ensuring it aligns seamlessly with the model's operational framework requires further development. The process of storing and retrieving relevant experiences from the Artificial Neuron must be refined to prevent performance bottlenecks.

\textbf{Quality Control}: Ensuring the quality and relevance of the feedback stored in the Artificial Neuron is crucial. Incorrect or misleading feedback could reinforce poor reasoning habits in the LLM, potentially degrading its performance over time.

\section{Conclusion}
This paper presents a pioneering approach to improving the reasoning abilities and decision-making accuracy of Large Language Models through the use of an external memory system, termed the Artificial Neuron. By allowing LLMs to learn from their past interactions, we facilitate a form of experiential learning that mimics human cognitive processes, enabling these models to perform complex problem-solving tasks with greater accuracy.

Future iterations of this methodology should focus on automating the feedback mechanisms. Developing algorithms that can automatically assess the effectiveness of an LLM’s responses and provide accurate corrections will enhance the scalability and efficiency of this approach.
- **Integration with User Feedback**: Implementing systems that allow LLMs to adjust their responses based on direct user feedback could provide a more dynamic learning environment. This could involve real-time adjustments to responses based on user interactions or satisfaction ratings.

We could explore the integration of the Artificial Neuron within a multi-agent framework that could provide insights into how shared external memories influence cooperative task-solving among multiple LLMs \cite{liang2023encouraging} \cite{qian2023communicative} \cite{woolley2010evidence} \cite{lazaridou2020multiagent} \cite{graesser2020emergent} \cite{lee2018emergent} \cite{wu2023large} \cite{rasal2024llm} \cite{cruz2024transforming} \cite{Musumeci2024LLMBM}. This could lead to advancements in how LLMs collaborate and learn from each other, potentially opening new avenues for developing more sophisticated AI systems.

By addressing these limitations and exploring these future directions, we can continue to refine the Artificial Neuron concept and expand its applicability across various domains, moving closer to creating LLMs that can operate with a level of autonomy and reasoning complexity that mirrors human intelligence.



\bibliographystyle{unsrt}
\bibliography{references}  

\end{document}